\documentclass[12pt, a4paper]{article}

\usepackage{units}
\usepackage{ucs}
\usepackage{hyperref}
\usepackage{graphicx}
\usepackage{times, mathptmx}
\usepackage{fancyhdr, xspace}

\begin{document}

%%line numbers, for referees
%\modulolinenumbers[1]
%\linenumbers

\title{HEAT -- a low energy enhancement of the Pierre Auger Observatory}

\author{
C. Meurer\footnote{RWTH Aachen University, III. Physikalisches Institut A, Aachen, Germany 
\newline
Correspondence to: C.~Meurer (christine.meurer@physik.rwth-aachen.de)}
\xspace and N. Scharf$^*$ \\ on behalf of 
the Pierre Auger Collaboration
\footnote{Observatorio Pierre Auger, Av. San Martin Norte 304, (5613) Malarg\"ue, Mendoza, Argentina  
%\newline
Full author list: \url{http://www.auger.org/archive/authors_2010_10.html}}
}

\date{19 May 2011}

%\runningtitle{HEAT -- a low energy enhancement of the Pierre Auger Observatory}
%\runningauthor{C.~Meurer and N.~Scharf}
%\correspondence{C.~Meurer\\ (christine.meurer@physik.rwth-aachen.de)}

\maketitle

\begin{abstract}
The High Elevation Auger Telescopes (HEAT) are three tiltable fluorescence
telescopes which represent a low energy enhancement of the fluorescence
telescope system of the southern site of the Pierre Auger Observatory in Argentina.

The Pierre Auger Observatory is a hybrid cosmic ray detector consisting of
24 fluorescence telescopes to measure the fluorescence light of extensive air
showers complemented by 1600 water Cherenkov detectors to determine the particle
densities at ground. In this configuration air showers with a primary energy
of $10^{18}$\,eV and above are investigated.

By lowering the energy threshold by approximately one order of magnitude down to
a primary energy of $10^{17}$\,eV, HEAT provides the possibility to study the cosmic ray energy spectrum
and mass composition in a very interesting energy range, where the transition from galactic to extragalactic
cosmic rays is expected to happen.

The installation of HEAT was finished in 2009 and data have been taken continuously
since September 2009. Within these proceedings the HEAT concept is presented.
First data already demonstrate the excellent complement of the standard Auger 
fluorescence telescopes
by HEAT.
\end{abstract}

\section{Exploring the nature of cosmic rays at the transition energy region}
\begin{figure}[h]
\vspace*{2mm}
\center
\includegraphics[width=9.5cm, viewport= 0 0 530 380,clip]{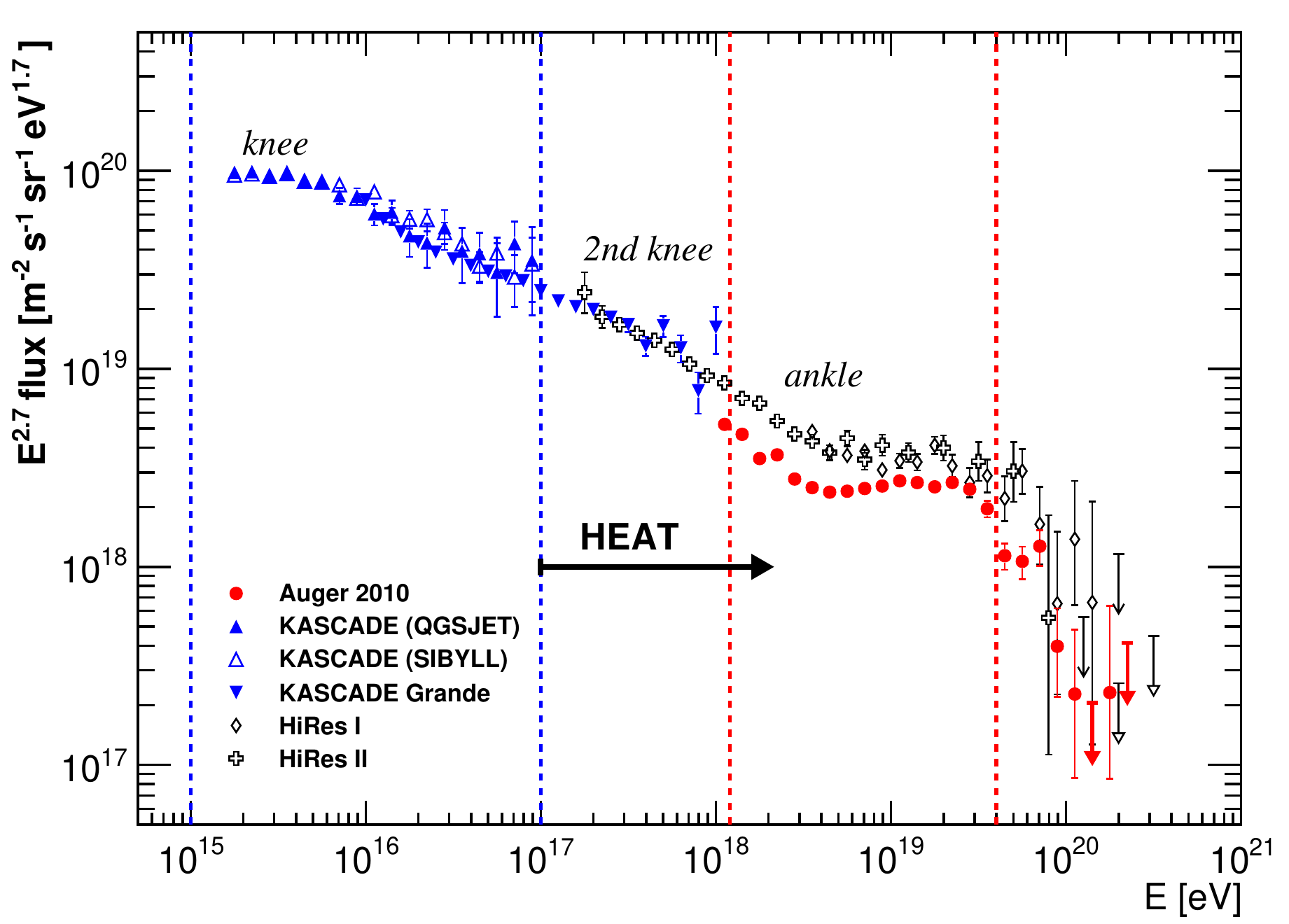}
\caption{\label{spectrum} Cosmic ray energy spectrum. The particle flux is scaled with $E^{ 2.7}$ to let
the three points of discontinuity ({knee}, {2nd knee} and {ankle}) become clearly visible.
The dashed lines mark the energy regions where the KASCADE (\unit{$10^{15}-10^{17}$}\,{eV}) and Auger
(\unit{$1.2\cdot10^{18}-4\cdot10^{19}$}\,{eV}) collaboration have published mass composition studies
\cite{kascade,augerXmax}.
The arrow starting from $10^{17}$\,eV indicates the energy range where HEAT has the possibility
to analyze the mass composition.
Data are taken from \cite{kascade,grande,hires,auger_energy2010}, only statistical error bars are shown.
}
\end{figure}

The energy range starting from $10^{17}$\,eV up to higher energies provides a
rich physics potential.
Two points of discontinuity in the cosmic ray energy spectrum known as {2nd knee} and {ankle}
are observed in this range (see e.g.\ \cite{bluemer_review2009,auger_energy2010}) and are visible
in the spectrum in Fig.~\ref{spectrum}. The transition from galactic to extragalactic cosmic rays is
expected to happen between these two features.
It is likely that different types of astrophysical sources are responsible for these two populations.
Consequently the composition of the cosmic rays also would be expected to change in this energy region.
Since there are so many exciting and not yet answered questions regarding the origin and nature of cosmic rays in
this part of the cosmic ray spectrum, it is desirable to have several independent measurements with different methods
of composition determination.

\newpage

One method to study the mass composition of cosmic rays is to measure the number of
electrons and muons at ground level as was done by KASCADE \cite{kascadeNIM} and other surface detector arrays.
The electron to muon ratio is sensitive to the mass and energy of the primary particle initiating the
extensive air shower.

Another method used by the Pierre Auger Observatory is to collect the fluorescence light
produced by the relaxation of nitrogen molecules in the earth's atmosphere which are excited by charged
particles of an extensive air shower. The atmospheric depth $X_{\mathrm{max}}$ is determined
at which the longitudinal shower development reaches its maximum in terms of energy deposit in the atmosphere
which is proportional to the number of secondary particles at this altitude.
The shower maximum $X_{\mathrm{max}}$ is an observable sensitive to the mass of the primary particle.
A proton induced shower has its maximum at lower altitude (larger atmospheric depth) than an iron induced shower of the
same energy. This behavior can be explained by the generalized Heitler model \cite{heitler1944,matthews2005}
and the superposition assumption for nuclear primary particles.

\section{Method of low energy enhancement}

Each of the standard fluorescence telescopes has a field of view of $30^{\circ} \times 30^{\circ}$ from 0\textdegree\ to
30\textdegree\ in elevation \cite{FD_NIM2010}. An extensive air shower initiated by a primary particle with an energy
lower than $10^{18}$\,eV emits less fluorescence light than showers with higher energy. Therefore, such a shower
can only be detected if it is close to a fluorescence telescope. Showers detected by the fluorescence telescopes will
only be used for physics analyses if the reconstructed shower maximum $X_{\mathrm{max}}$ is in the field of view
of the telescopes. In order to detect the shower maximum of showers close to the telescopes, they must have a field of
view that extends to higher elevations. This is realized by the additional field of view of HEAT from 30\textdegree\ to
60\textdegree\ in elevation.

\begin{figure}[h]
\vspace*{2mm}
\center 
\includegraphics[width=12cm, viewport=  30 360 600 670,clip]{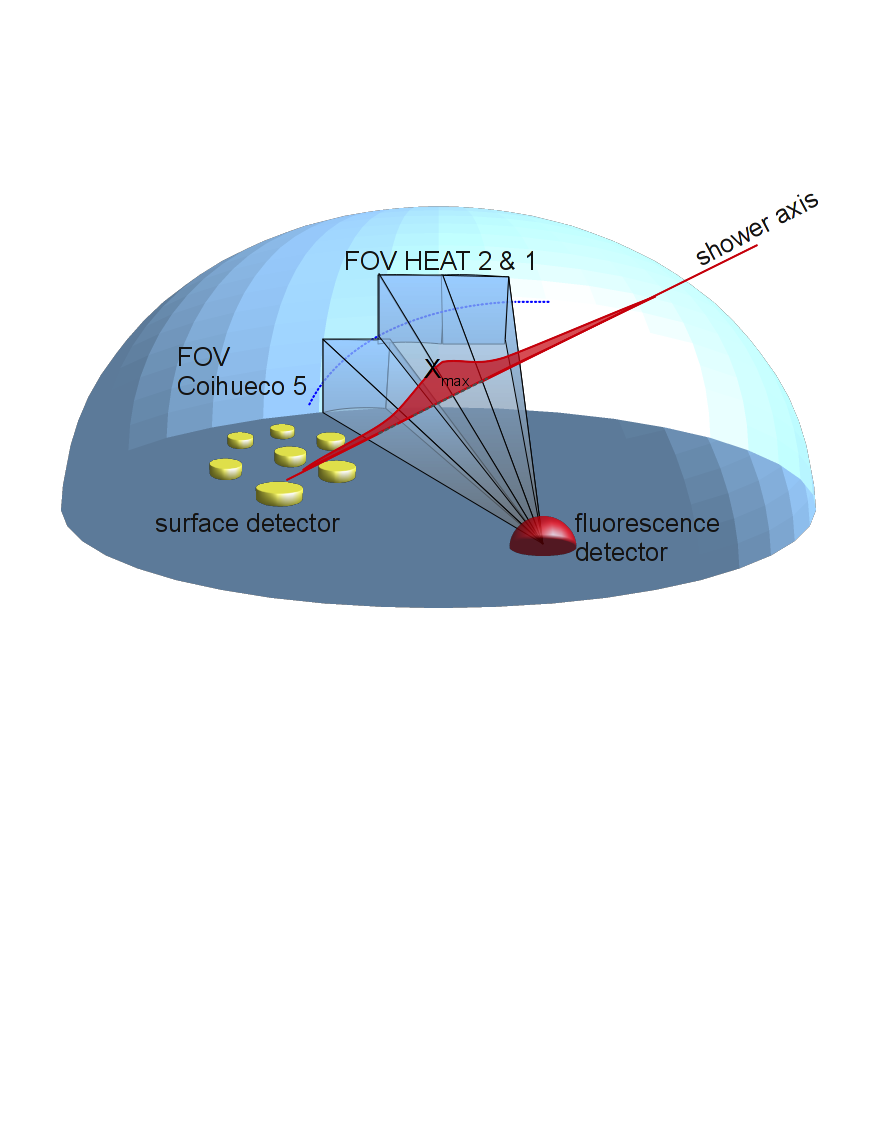}
\caption{\label{HEAT_FOV}  Schematic illustration of the shower detection method by a standard fluorescence telescope in 
combination with two HEAT telescopes. The field of view (FOV) of a standard fluorescence telescope (Coihueco 5) 
and two HEAT telescopes (HEAT 2 \& 1) are depicted.}
\end{figure}

Figure~\ref{HEAT_FOV} shows a schematic illustration of the shower detection method by a standard fluorescence telescope in
combination with two HEAT telescopes. The shower maximum $X_{\mathrm{max}}$ of the indicated shower is seen in
the field of view of one of the HEAT telescopes. A shower with such a geometry would not pass the selection criteria for
physics analyses if it was detected by the standard fluorescence telescope only.

%\item A shower is only used for physics analysis if the reconstructed
%shower maximum $X_{\mathrm{max}}$ is in the field of view of the fluorescence telescopes.
%\item Low energy showers emit less fluorescence light and therefore they can only be detected closer
%to the telescope.
%\item In order to detect showers closer by the telescopes, they have to overview a field of view at
%higher elevation to see the shower maximum.
%\item This geometric bias is reduced by the additional field of view of HEAT at higher elevation.

\section{High Elevation Auger Telescopes}

\begin{figure}
\center
\includegraphics[width=8.3cm]{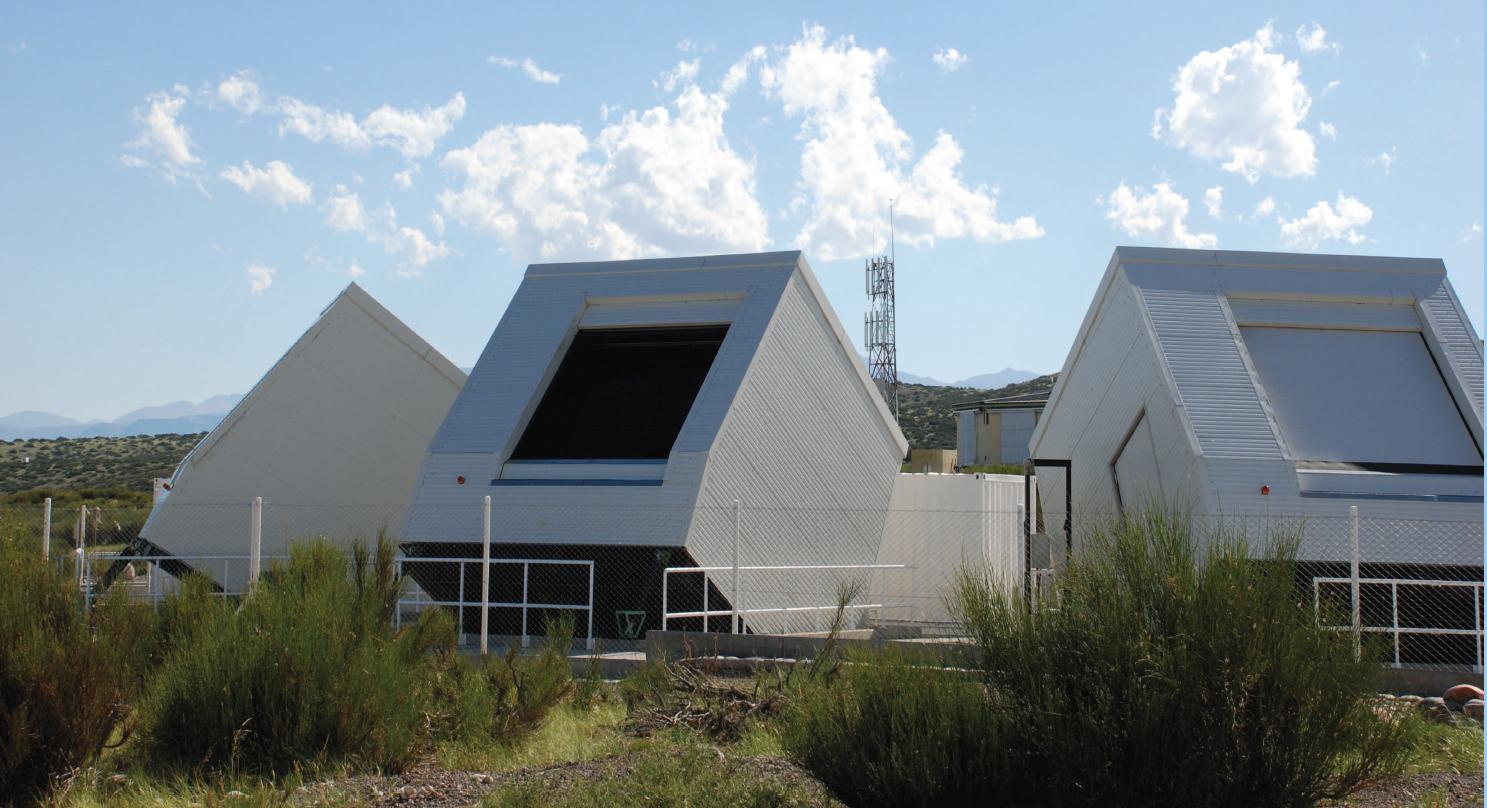}
\caption{\label{HEAT_Foto} Photo of the three HEAT buildings in tilted mode.}
\end{figure}

%HEAT site
The three tiltable HEAT buildings (see the photo in Fig.~\ref{HEAT_Foto}) are situated close to the existing
fluorescence building Coihueco housing six standard fluorescence telescopes. This configuration enables to measure
showers in the way indicated in Fig.~\ref{HEAT_FOV}.

%optics
HEAT uses the same Schmidt optics as implemented in the standard fluorescence telescopes.
For details of the construction of the Auger fluorescence telescopes see \cite{FD_NIM2010}.
%tilting & monitoring
The tilt hydraulic is driven by a commercial electric motor. For installation, calibration and also for comparison
with data taken by Coihueco, HEAT can be operated in untilted mode as well.
Inclination sensors at mirror, camera and aperture box and distance sensors between camera and mirror system are used
to monitor the camera position in the optical aperture of the telescopes in a tilted position in order to guarantee the
stability of the camera position.
%DAQ & trigger rate
The data acquisition (DAQ) electronics of HEAT has an increased sampling rate of \unit{20}{MHz} instead of \unit{10}{MHz}
for the standard fluorescence telescopes. This is required to sample signals of showers closer to the telescopes,
since they appear in the field of view of the telescopes for shorter time.
%has a higher time resolution (\unit{50}{ns} instead of \unit{100}{ns})
The readout electronics is able to handle the increased event rate expected at lower energy of the primary particle.
In addition, the HEAT DAQ system is a prototype for the fluorescence telescopes at the northern site of
the Pierre Auger Observatory.
For more information on the HEAT design see \cite{kleifges_icrc2009}.

%\begin{itemize}
%\item Three additional tiltable fluorescence telescopes close to existing fluorescence telescopes (Coihueco),
%\item Auger low energy enhancement HEAT
%\item standard FD telescope, Schmidt optics, see NIM FD paper
%\item differences: DAQ electronics, higher time resolution: instead of 100ns 50ns (10MHz to 20MHz),
%upgraded electronics to handle increased trigger rates
%\item Lowering energy threshold of Auger by one order of magnitude down to $10^{17}$\,eV by extending
%the field of view in elevation: 0\degree- 30\degree (standard Auger telescopes) + 30\degree- 60\degree (HEAT)
%\item hydraulic of tilting (electric motor)
%\item Monitoring of camera position in optical aperture of telescope in tilted position
%to guarantee stability of camera position using inclination and distance sensors
%\item calibration like FD
%\item HEAT as fifth eye in trigger logic
%\end{itemize}

\section{HEAT data taking}

The installation of HEAT was finished in 2009 and data have been taken  continuously since September 2009.
The duty cycle of the new telescope system is identical to the existing standard fluorescence telescopes,
about \unit{13}{\%} due to the necessity of clear, moonless nighttime conditions.

\begin{figure}
\vspace*{2mm}
\center
\includegraphics[width=9cm]{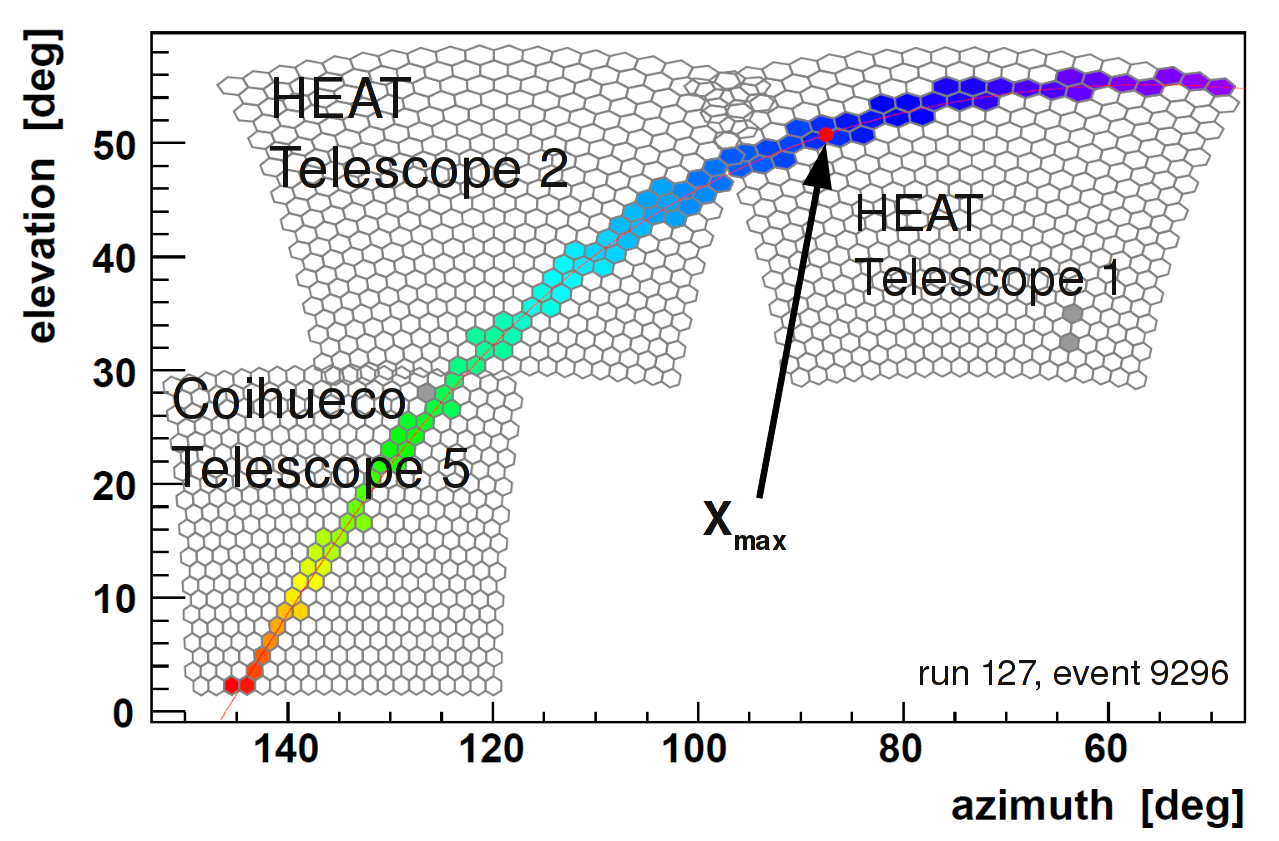}
\caption{\label{EventDisplay} Event display of two HEAT and one Coihueco cameras. The photomultiplier pixels record
the fluorescence light of an extensive air shower starting from the upper right corner to the lower left.
The position of the shower maximum  $X_{\mathrm{max}}$ is seen by telescope~1 of HEAT.}
\end{figure}

In Fig.~\ref{EventDisplay} a recorded shower with a similar geometry like that in Fig.~\ref{HEAT_FOV}
is shown in the event display of two HEAT cameras and one Coihueco camera. The position of the shower maximum lies
in the field of view of the telescope~1 of HEAT and, therefore, this shower would not pass the selection criteria for
physics analyses without the HEAT measurement.

\begin{figure}
\vspace*{2mm}
\center
\includegraphics[width=9cm]{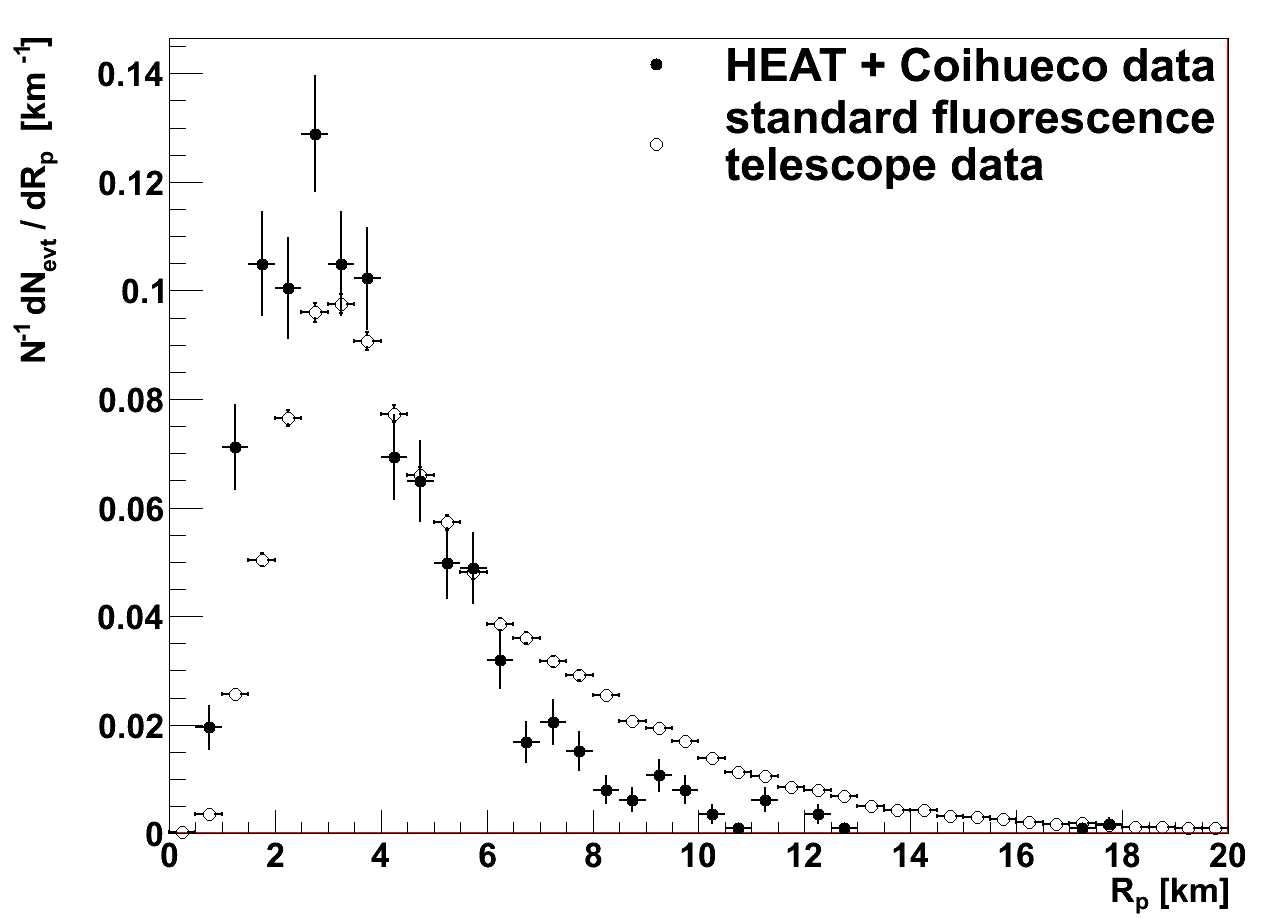}
\caption{\label{minDist} Minimal distance $R_p$ from the detector to the shower axis. The distribution for
the combination of HEAT and Coihueco and for standard fluorescence telescopes are normalized by the total number
of registered showers in each detector. Only showers with $X_{\mathrm{max}}$ in the field of view
are considered.}
\end{figure}

Looking at the first HEAT data the additional benefit of this Auger enhancement becomes clearly visible.
More extensive air showers closer to the fluorescence telescopes are recorded by the combination of HEAT and
Coihueco ($0^{\circ} - 60^{\circ}$) compared to the standard fluorescence telescopes \mbox{($0^{\circ} - 30^{\circ}$)}.
This can be seen in Fig.~\ref{minDist} where the distribution of the minimal distance $R_p$
from the detector to the shower axis is shown for HEAT+Coihueco and for standard fluorescence data.
More showers with a smaller $R_p$ are collected by HEAT+Coihueco which increases the statistics of showers with
a lower primary energy.

\enlargethispage{0.7cm}

Another geometrical bias of the standard fluorescence telescopes is reduced by the additional detected showers.
Having a field of view in elevation up to 30\textdegree and considering the quality criterion that the shower maximum
has to be in the field of view, it is more likely to measure showers crossing the field of view from the front
than from the back.
This behavior is still visible in the showers recorded by HEAT+Coihueco but much less pronounced,
i.e.\ the combination of HEAT and Coihueco is less dependent on whether a shower crosses the field of view from the front or back.

\begin{figure}
\vspace*{2mm}
\center
\includegraphics[width=9cm]{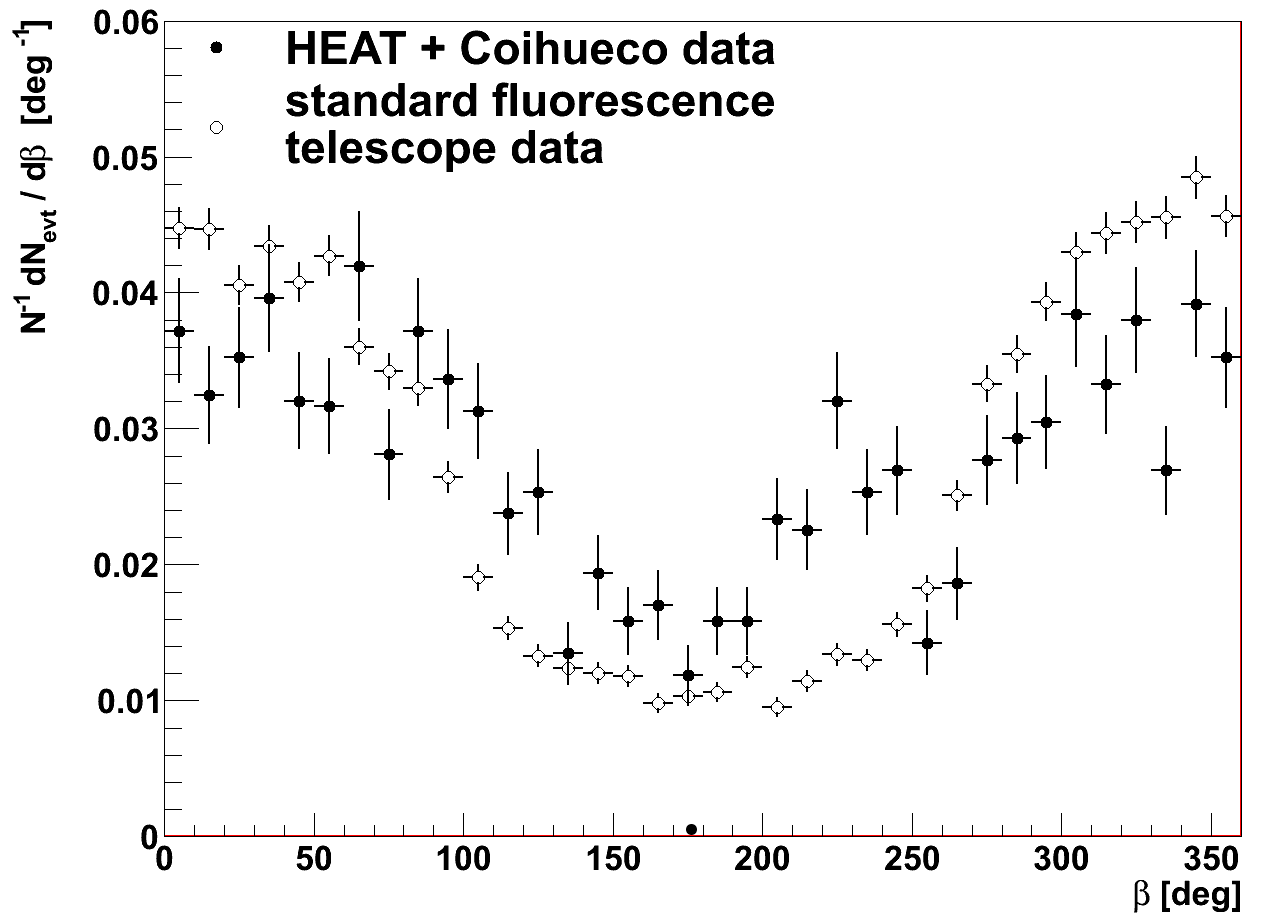}
\caption{\label{azimuth} Distribution of angle $\beta$, the difference of the azimuth angle of a shower and the
azimuth angle of the mean line of sight of the telescopes. $\beta=0^{\circ}$ or $360^{\circ}$ corresponds to the front
of the detector, $\beta=180^{\circ}$ to its back. The distribution for the combination of HEAT and Coihueco and for
standard fluorescence telescopes are normalized by the total number of registered showers in each detector.
Only showers with $X_{\mathrm{max}}$ in the field of view are considered.}
\end{figure}

In Fig.~\ref{azimuth} the distribution of the angle $\beta$ is plotted for HEAT+Coihueco and for standard fluorescence
telescopes, where $\beta$ is defined as the difference of the azimuth angle of the shower and the azimuth angle of the
mean line of sight of the telescopes. A shower with $\beta=180^{\circ}$ reaches the telescopes from the back, a shower with
$\beta=0^{\circ}$ or $360^{\circ}$ from the front. The distribution of $\beta$ is considerably flatter for HEAT+Coihueco
than for the standard fluorescence telescopes. Hence, more showers reaching the telescopes from the back side are
measured and provide the possibility to measure showers with lower primary energy.

%\begin{itemize}
%\item Installation of HEAT finished in 2009
%\item Data taking since September 2009
%\item Duty cycle of HEAT: 15\% (only in moonless nights),
%same duty cycle as standard Auger telescopes,
%\item HEAT trigger rate: ~1 trigger/min,
%10x higher than trigger rate of standard Auger telescopes,
%as assumed from steeply decreasing energy spectrum
%\item event display
%\item HEAT detects more showers closer to the detector and less at larger distance
%\item HEAT is less dependent on whether a shower crosses its field of view from front or back.
%\end{itemize}

\section{Conclusions and outlook}

The low energy enhancement HEAT of the Pierre Auger Observatory has operated well since its installation in 2009.
The first data demonstrate the complementary performance of HEAT. Extensive air showers closer to the detector and such
showers reaching the field of view of the telescopes from the back side are measured by the combination of HEAT
and Coihueco with high statistics and, therefore, provide the possibility to study showers with lower primary energy
down to at least $10^{17}$\,eV.
This gives the opportunity to analyze the cosmic ray energy spectrum and the mass composition in a very interesting
energy range where the transition from galactic to extragalactic cosmic rays is expected to happen.
First physics results are planned for 2011.

%In the area where HEAT is located also other Auger enhancements operating in similar energy regions are situated: \\

Other Auger enhancements operating in energy regions similar to HEAT are also located in the same general area: \\
(a) AMIGA, {Auger Muons and Infill for the Ground Array}, consisting of an underground muon detector and the
 {infill} array, an additional field of surface detectors arranged on a grid with smaller spacing than the
standard surface detector array. \\
(b) AERA, the {Auger Engineering Radio Array}, an array of antennas to measure the radio signals of
extensive air showers mainly produced by the synchrotron effect of charged particles in the geomagnetic field.

A long-term goal is to combine step by step these Auger enhancements for a multi detector analysis.
In a first step HEAT and the {infill} will be used for a hybrid shower reconstruction like it is done in the
standard analysis method of fluorescence telescopes and surface detectors.

{\small\noindent
Edited by: T.~Suomijarvi\\
Reviewed by: two anonymous referees}


\begin{thebibliography}{}

\bibitem[Abbasi et al., 2008]{hires}
Abbasi, R.~U., Abu-Zayyad,T., Allen, M.\ et al.\ [HiRes]:
First Observation of the Greisen-Zatsepin-Kuzmin Suppression,
Phys.\ Rev.\ Lett., 100, 101101-1 - 101101-5, 2008.
%HiRes I, II

\bibitem[Abraham et al., 2010a]{auger_energy2010}
Abraham, J., Abreu, P., Aglietta, M.\ et al.\ [The Pierre Auger Collaboration]:
Measurement of the energy spectrum of cosmic rays above $10^{18}$ eV using the Pierre Auger Observatory,
Phys.\ Lett.\ B, 685, 239-246, 2010a.
%Auger, J. Abraham et al.
%F. Schüssler, Proc. of 31th Int. Cosmic Ray Conf. 2009

\bibitem[Abraham et al.(2010b)]{augerXmax}
Abraham, J., Abreu, P., Aglietta, M.\ et al.\ [The Pierre Auger Collaboration]:
Measurement of the Depth of Maximum of Extensive Air Showers above \unit{$10^{18}$}{eV},
Phys.\ Rev.\ Lett.\ 104, 091101-1 - 091101-7, 2010b.

\bibitem[Abraham et al.(2010c)]{FD_NIM2010}
Abraham, J., Abreu, P., Aglietta, M.\ et al.\ [The Pierre Auger Collaboration]:
The Fluorescence Detector of the Pierre Auger Observatory,
Nucl.\ Instrum.\ Meth.\ A 620, 227-251, 2010c.
%Auger, J. Abraham et al.

\bibitem[Antoni et al.(2003)]{kascadeNIM}
Antoni, T., Apel, W.~D., Badea, F.\ et al.\ [KASCADE]:
The Cosmic-Ray Experiment KASCADE,
Nucl.\ Instrum.\ Meth.\ A, 513, 490-510, 2003.
%KASCADE NIM paper

\bibitem[Antoni et al.(2005)]{kascade}
Antoni, T., Apel, W.~D., Badea, A.~F.\ et al.\ [KASCADE]:
KASCADE measurements of energy spectra for elemental groups
of cosmic rays: Results and open problems,
Astropart.\ Phys.\ 24, 1-25, 2005.
%KASCADE

\bibitem[Bertaina(2009)]{grande}
Bertaina, M.\ for the KASCADE Grande Coll.:
The all particle energy spectrum of KASCADE-Grande in the
energy region \unit{$10^{16} - 10^{18}$}{eV} by means of
the $N_{\textrm{ch}} - N_\mu$ technique,
Proc.\ 31th Int.\ Cosmic Ray Conf., 2009. %15-18
%KASCADE Grande

\bibitem[Bl\"umer et al.(2009)]{bluemer_review2009}
Bl\"umer, J., Engel, R. and H\"orandel J.:
Cosmic rays from the knee to the highest energies,
Prog.\ in Part.\ and Nucl.\ Phys.\ 63, 293-338, 2009.

\bibitem[Heitler(1944)]{heitler1944}
Heitler, W.:
The Quantum Theory of Radiation,
Oxford University Press, 3rd edition, 1954.

\bibitem[Kleifges(2009)]{kleifges_icrc2009}
Kleifges, M. for the Pierre Auger Coll.:
Extension of the Pierre Auger Observatory using high-elevation
fluorescence telescopes (HEAT),
Proc.\ 31th Int.\ Cosmic Ray Conf., 2009.
%HEAT ICRC 2009 proceeding

\bibitem[Matthews(2005)]{matthews2005}
Matthews, J.:
A Heitler model of extensive air showers,
Astropart.\ Phys.\ 22, 387-397, 2005.



\end{thebibliography}
\end{document}